# The effect of exceptionally high fluorine doping on the anisotropy of single crystalline $SmFeAsO_{1-x}F_x$

Masaya Fujioka[1,a)], Saleem J. Denholme[1], Masashi Tanaka[1] Hiroyuki Takeya[1], Takahide Yamaguchi[1], and Yoshihiko Takano[1,2]

[1] *National Institute for Materials Science, 1-2-1 Sengen, Tsukuba, Ibaraki 305-0047, Japan*

[2] *University of Tsukuba, 1-1-1Tennodai, Tsukuba, Ibaraki 305-0001, Japan*

Abstract

We succeeded in growing single crystalline $SmFeAsO_{1-x}F_x$ with exceptionally high fluorine concentration by using a CsCl flux method. Comparing to conventional flux methods, this method can introduce about double the amount of fluorine into the oxygen site. The obtained single crystal shows the highest superconducting transition temperature ($T_c$ = 57.5 K) in single crystalline iron pnictides. In addition, the residual resistivity ratio (RRR) is almost three times as large as that of previously reported single crystals. This suggests that our single crystals are suitable for investigation of the intrinsic superconducting properties since they have few defects and impurities. Using both the Werthamer-Helfand-Hohenberg model and the effective mass model, we demonstrated that a higher fluorine concentration suppresses the anisotropic superconductivity of $SmFeAsO_{1-x}F_x$.

The superconducting oxypnictides (1111 system) are based on alternating structures of conducting and blocking layers [1]. To induce superconductivity, it is necessary to introduce electron carriers into the conducting layers in this system [2-4]. A partial fluorine substitution is one of the typical methods for carrier doping. However, the over-doped regions for fluorine doping have not been reported; superconducting transition temperature ($T_c$) continues to increase with increasing fluorine concentration [5]. Recently, some groups reported that low temperature sintering enables the synthesis of SmFeAsO$_{1-x}$F$_x$ with a higher $T_c$ when compared to traditional sintering [6-9]. Actually, it is difficult to obtain a $T_c$ above 55 K by traditional sintering method. However, the low temperature sintering makes it possible. The current highest $T_c$ (58.1 K) in iron pnictides is obtained by this method [10].

Figure 1 (a) shows fluorine content dependence of the $a$ lattice constant. The black open circles denote the reported results for polycrystalline SmFeAsO$_{1-x}$F$_x$ prepared by low temperature sintering. $T_c$ drastically increases up to around 55 K, and the $a$ lattice parameter rapidly decreases until $x = 0.15$. The $a$ lattice parameter's tendency to decrease clearly changes at $x = 0.15$, which is due to the difficulty of further fluorine doping. Therefore, we have established that the critical point to separate the low fluorine (LF) and high fluorine (HF) regions is at $x = 0.15$ in this research.

According to the previous reports, the $T_c$ is strongly related to the $a$ lattice constant [8,9]. Figure 1 (b) shows the relationship between $T_c$ and $a$ lattice constant obtained from various reports about both single crystalline and polycrystalline SmFeAsO$_{1-x}$F$_x$. Although a large increase in $T_c$ is not observed in the HF region, the *a* lattice constant continues to shrink due to further fluorine doping above $x = 0.15$. Up until now, the reported $T_c$ of a single crystal with fluorine doping has been below 53 K as shown in the area enclosed by the red dotted line in figure 1 (a) [11-16]; all reported single crystals are within the LF region below $x = 0.12$. Namely fluorine substitution into single crystalline $SmFeAsO_{1-x}F_x$ is more difficult than into the polycrystalline equivalent. However, the region of much higher fluorine concentration for a single crystal should be investigated to reveal the intrinsic properties of the F-doped 1111 system. In this research, we succeeded in preparing single crystals of $SmFeAsO_{1-x}F_x$ with exceptionally high fluorine concentrations as shown in figure 1 (a) and (b).

For the sample preparation, a quartz ampule was used to prevent the leak of fluorine during the growth of the single crystals. CsCl was adopted as a flux, because it does not react with a quartz even over 1000 °C [17]. The starting materials were prepared with the same method reported in ref 10. The nominal fluorine concentrations were adopted as $x = 0.0, 0.2, 0.3, 0.4$ and $0.5$. A 1 : 6 ratio of starting materials and CsCl were weighted and ground together. Then, they were sintered at 950 °C for 5 h after sealing a quartz tube and slowly cooled at a rate of 0.5 °C / h down to 700 °C. The size of the obtained single crystals was around 10 μm. Therefore, it is necessary to use the fine processing technology for resistivity measurements. First, one single crystal of $SmFeAsO_{1-x}F_x$ was positioned on a MgO substrate. Then, it was connected with tungsten electrodes using focused ion beam (FIB) and formed into a rectangle as shown in the inset of figure 2. Resistivity measurements were performed under magnetic fields (0, 1, 3, 5, 7, 9, 11, 13, 15 T) with directions both parallel to the *ab* plane

($H^{//ab}$) and c axis ($H^{//c}$). The angle dependence of resistivity was also measured at various temperatures and magnetic fields for the effective mass method[18-20].

Figure 2 shows the temperature dependence of resistivity for two single crystals within LF and HF regions, which are called LFs and HFs respectively. The $T_c$ of LFs and HFs were obtained at 53.5 K and 57.5 K respectively. At present, 57.5 K is the highest superconducting transition temperature in the single crystalline iron-based superconductors. Also their resistivity lineally decreases with decreasing temperature. The residual resistivity ratio (RRR = $\rho(300\ K)/\rho(T_c)$) are estimated to be 11.6 and 8.4 for LFs and HFs respectively. While previously reported RRR of single crystals are only about from 2 to 4 [13, 21]. This means that the samples obtained by CsCl flux method include far fewer defects and impurities. Therefore our single crystals should be sutable to investigate intrinsic superconducting properties of this system.

The temperature dependence of resistivity under magnetic field up to 15 T was measured for each sample as shown in figure 3. In the case of the superconductor with a layered structure, the suppression of superconductivity under $H^{//ab}$ is smaller than that under $H^{//c}$. Our samples also show the typical behavior of layered superconductors. Figure 4 shows magnetic field versus temperature in terms of $H_{c2}$ and irreversibility field ($H_{irr}$). 90 % of resistivity at $T_c^{onset}$ is adopted as a criterion for the estimation of $H_{c2}$ ,and $H_{irr}$ is determined at $\rho = 0$. Although the maximum applied magnetic field (15 T) is too small to estimate the $H_{c2}$ of this material, superconducting anisotropy is calculated from the initial gradient ($dH_{c2}/dT$) by the following formula of the Werthamer-Helfand-Hohenberg

(WHH) model; $\gamma = (\partial H_{c2}^{//ab}(T)/\partial T |_{T \to T_c} T_c) / (\partial H_{c2}^{//c}(T)/\partial T |_{T \to T_c} T_c)$. Estimated initial gradients are described in figure 4. From those values, the calculated $\gamma$ of LFs and HFs are 5.5 and 2.8 respectively. These results mean that a larger amount of fluorine doping suppresses the anisotropic superconducting properties.

The effective mass model was also applied to estimate the anisotropy of single crystals. In this method, the angle dependence of resistivity in the flux liquid state is needed to evaluate the reduced field ($H_{red}$), which is defined by $H_{red} = H(\sin^2\theta(R) + \gamma^{-2}\cos^2\theta(R))^{1/2}$, where $\theta$ is the angle between the *ab*-plane and the direction of magnetic field, and $R$ is resistance. The $\gamma$ is decided from the best scaling for the relationship between $R$ and $H_{red}$ [18-20]. This model is supposed to be applied only in the vortex liquid state. However, the step like behavior of resistivity under $H^{//c}$ observed in figure 3 may originate from a transition between the vortex liquid state and the pinned vortex liquid state [22]. Therefore, except for the data within the pinned liquid like state, only the data in limited magnetic fields and temperatures are shown in figure 5. A good scaling could be observed by taking $\gamma$ = 5.5, 7 and 10 at 48.2, 50.5 and 52.7 K respectively for LFs. In the same way, the $\gamma$ of HFs were estimated to be 3.5, 5.5 and 6 at 50, 52.5 and 54.6 K respectively. From the effective mass model, it is also demonstrated that a high level of fluorine doping results in a lower $\gamma$.

In summary, we succeeded in growing the single crystalline SmFeAsO with an exceptionally high fluorine concentration. The CsCl flux method using a quartz ampule is more effective to increase fluorine concentration than conventional flux method. The obtained $T_c$ (57.5 K) in this study shows the current-record for single crystalline iron-pnictides. In addition, the RRR is around three times larger than that of reported single crystals. This suggests that our single crystals are suitable to investigate the intrinsic superconducting properties due to their few defects and impurities, which are expected from large RRR. Using both the WHH model and the effective mass model, we demonstrated that a high level of fluorine doping induces a low anisotropy. As mentioned above, the over-doped regions for fluorine doping have not yet been observed in the 1111 system. Therefore, if further fluorine can be introduced into $SmFeAsO_{1-x}F_x$, it is expected to result in much higher $T_c$ and lower $\gamma$. Such a high level of fluorine doping for this system is one of the potential ways to develop superconducting applications.

This work was supported in part by the Japan Science and Technology Agency through Strategic International Collaborative Research Program (SICORP-EU-Japan) and Advanced Low Carbon Technology R&D Program (ALCA) of the Japan Science and Technology Agency.

Caption

Figure 1.

(a): The fluorine content versus $a$ lattice parameter. (b): The $a$ lattice parameter versus $T_c$. Open circles denote the reported results about polycrystalline $SmFeAsO_{1-x}F_x$ prepared by a low temperature sintering. Closed circles also denote the reported results about single crystalline $SmFeAsO_{1-x}F_x$. Red closed circles were obtained from this work. The actual fluorine concentrations of single crystals were determined by electron probe micro analyzer.

Figure 2.

Resistivity versus temperature for LFs and HFs. Upper inset shows the expanded view near $T_c$ of HFsc. Black lines are fitted lines for an estimation of $T_c$. Down inset shows the SEM image of the sample after the FIB process.

Figure 3.

Resistivity versus temperature for LFs and HFs under magnetic field with the different directions, which are parallel with ab plane ($H^{//ab}$) and c axis ($H^{//c}$) respectively.

Figure 4.

Upper magnetic field versus temperature for LFs and HFs. Solid circles denote the $H_{c2}^{//ab}$ and $H_{c2}^{//c}$. Open circles denote the $H_{irr}^{//ab}$ and $H_{irr}^{//c}$.

Figure 5.

The $R - H_{red}$ curves for LFs and HFs. The estimated anisotropy ($\gamma$) and applied magnetic fields are described for each measured temperature.

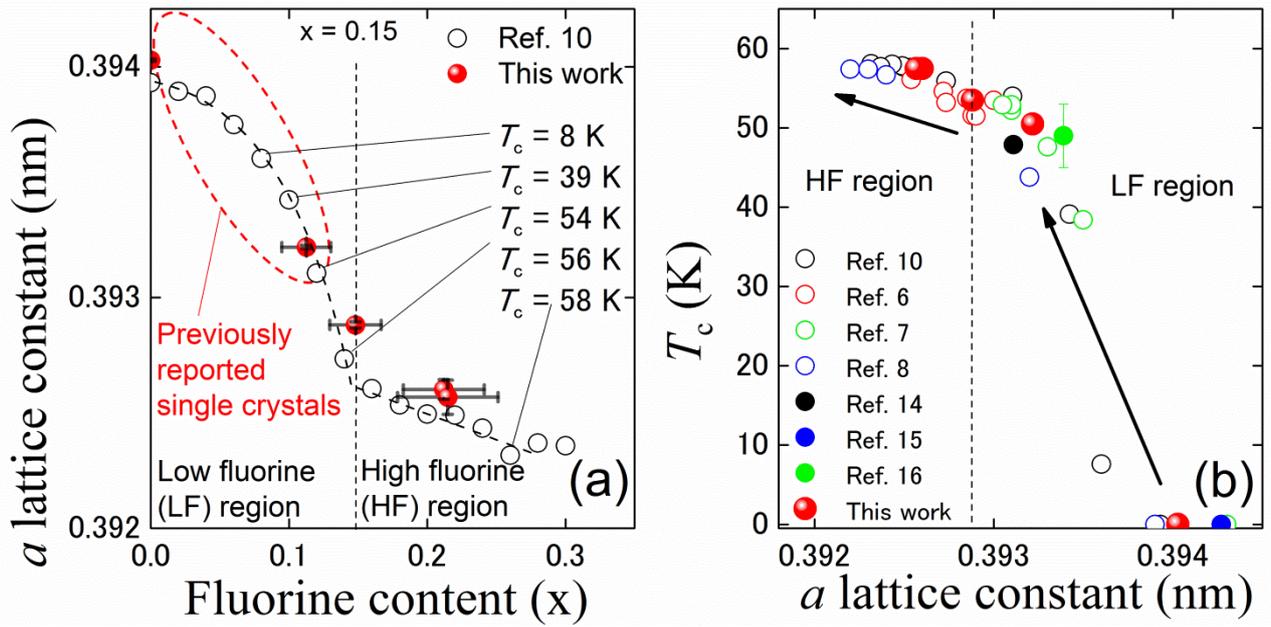

FIG.1

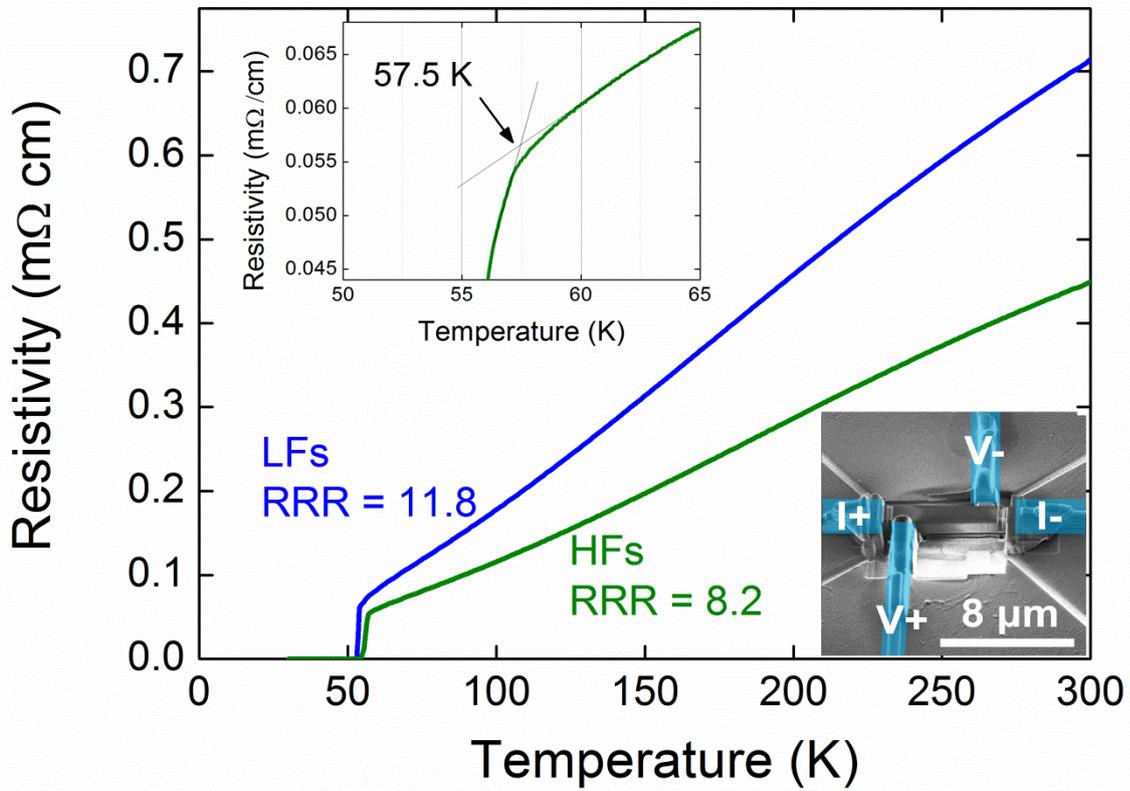

FIG.2

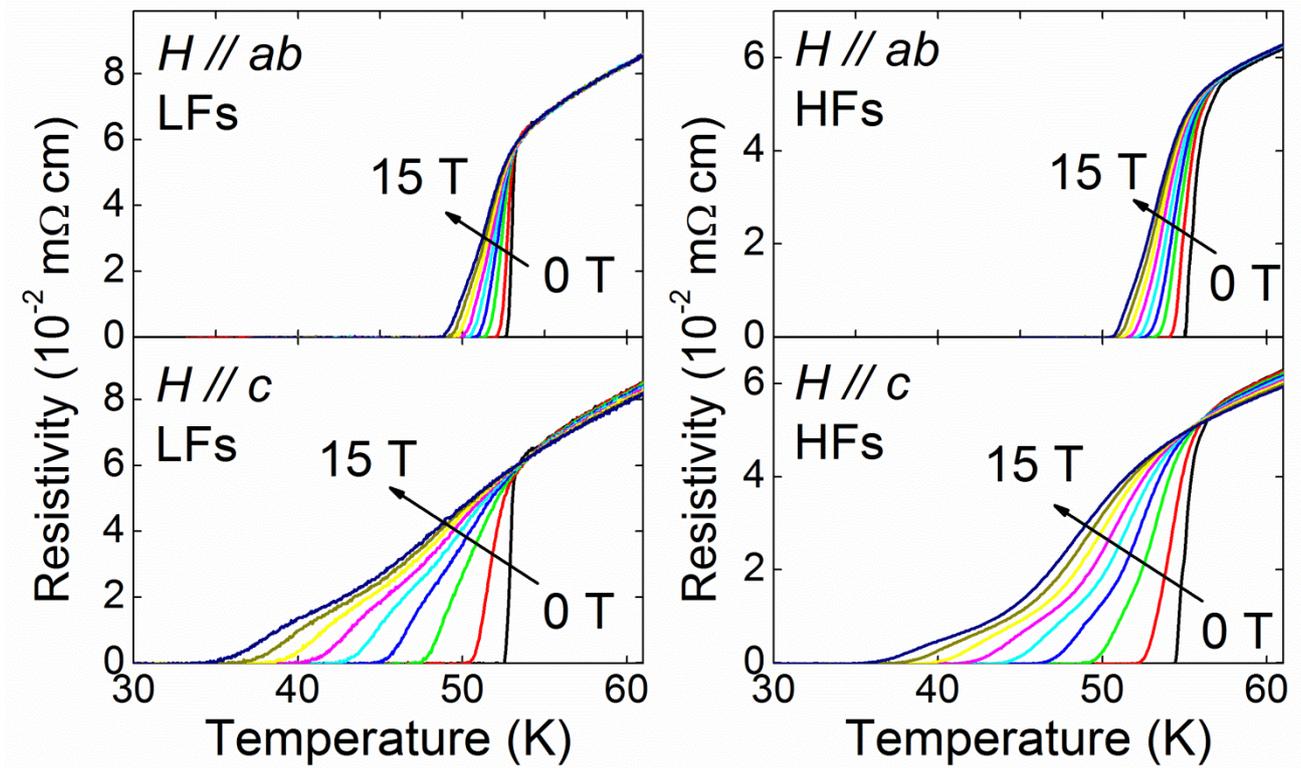

FIG.3

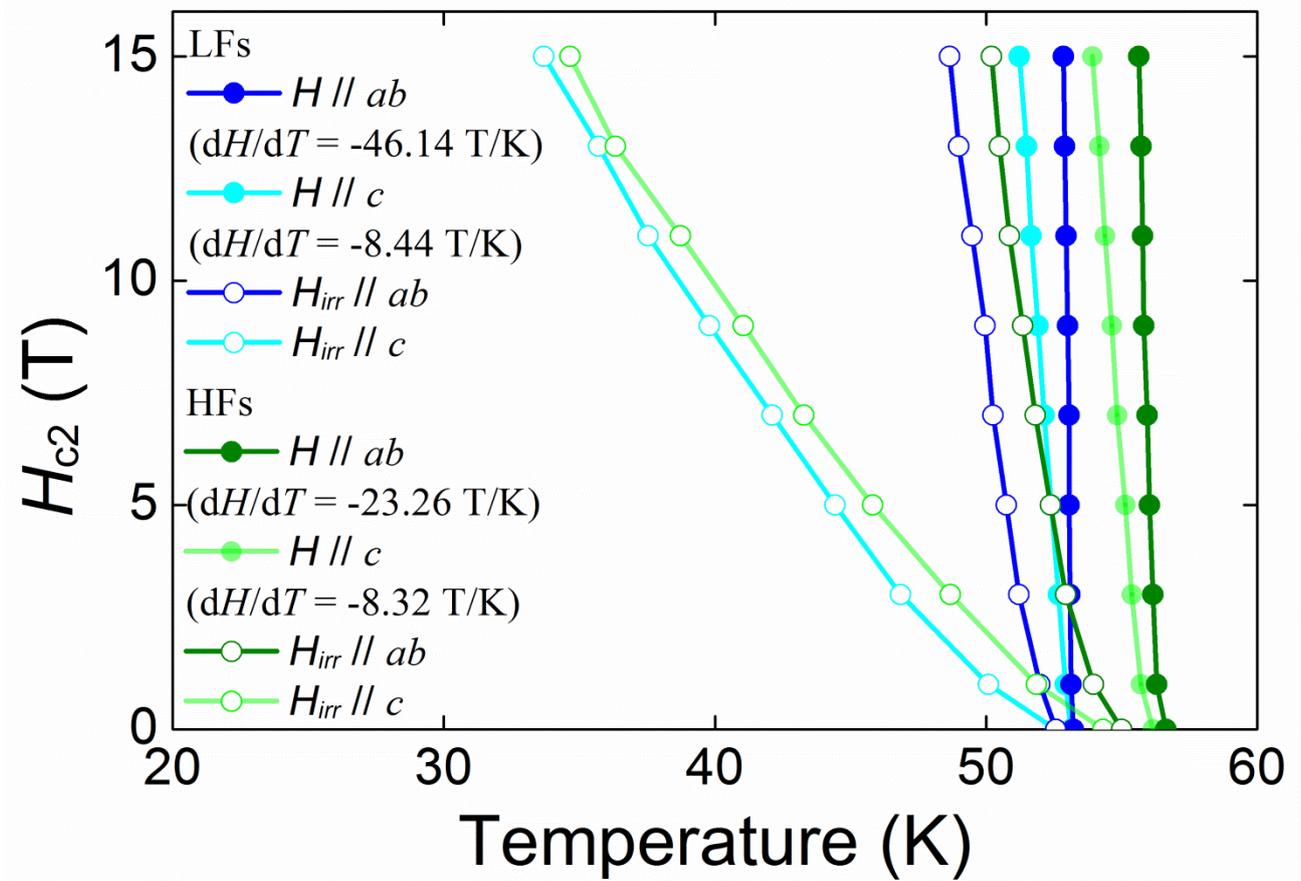

FIG. 4

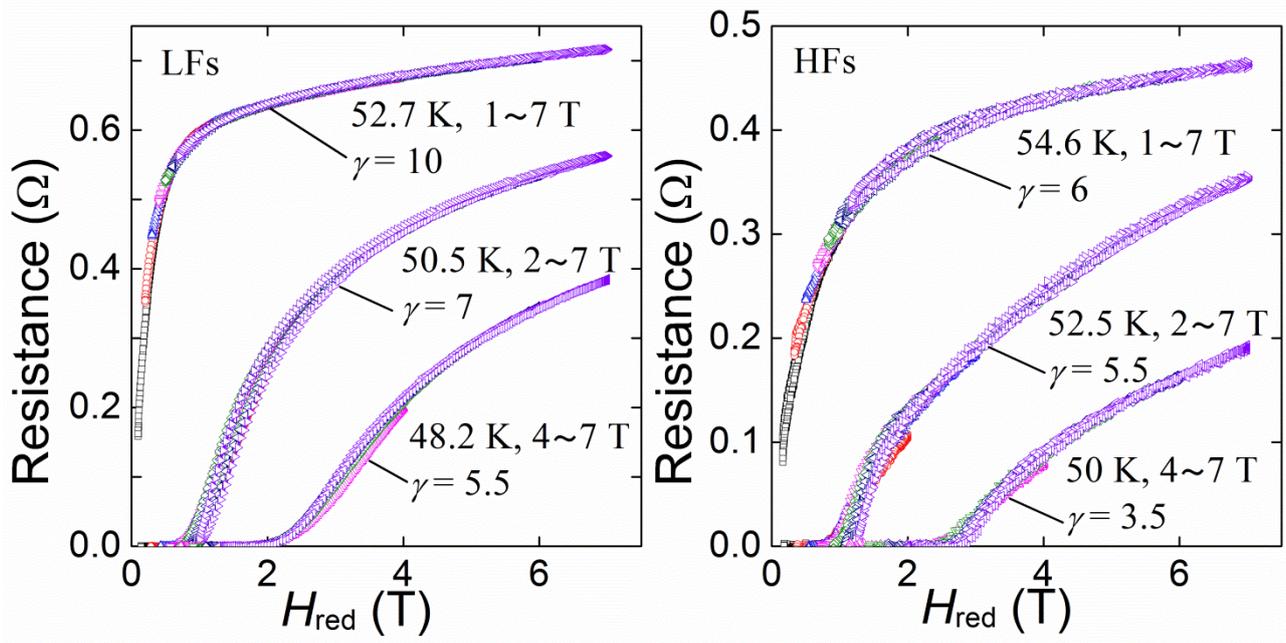

FIG.5